\documentclass[aip,jcp, preprint,cites,graphicx]{revtex4-1}
\usepackage{epsfig}
\usepackage{amsmath}
\usepackage{amssymb}
\usepackage{color}
\usepackage{setspace}
\usepackage{comment}
\usepackage{subfig}
\usepackage{indentfirst}

\begin{document}

\title{ÒExciton coherence lifetimes from electronic structure}

\author{John A. Parkhill} 
\email{john.parkhill@gmail.com}
\author{Alan Aspuru-Guzik} 
\email{aspuru@chemistry.harvard.edu}
\affiliation{
Department of Chemistry and Chemical Biology,
Harvard University, 
12 Oxford St. 
Cambridge, MA 02138, USA
}
\author{David Tempel} 
\affiliation{
Department of Physics,
Harvard University, 
17 Oxford St. 
Cambridge, MA 02138, USA
}

\begin{abstract}   
	We model the coherent energy transfer of an electronic excitation within covalently linked aromatic homodimers from first-principles, to answer whether the usual models of the bath calculated via detailed electronic structure calculations can reproduce the key dynamics. For these systems the timescales of coherent transport are experimentally known from time-dependent polarization anisotropy measurements, and so we can directly assess the whether current techniques might be predictive for this phenomenon. Two choices of electronic basis states are investigated, and their relative merits discussed regarding the predictions of the perturbative model. The coupling of the electronic degrees of freedom to the nuclear degrees of freedom is calculated rather than assumed, and the fluorescence anisotropy decay is directly reproduced.  Surprisingly we find that although TDDFT absolute energies are routinely in error by orders of magnitude more than the coupling energy, the coherent transport properties of these dimers can be semi-quantitatively reproduced from first-principles. The directions which must be pursued to yield predictive and reliable prediction of coherent transport are suggested. 
	
{\bf Keywords}: TDDFT, CRET, Coherence, electron-phonon coupling
\end{abstract}
 
\maketitle

\clearpage

\section{Introduction}
\begin{figure}
\begin{center}
\resizebox{140mm}{!}{\includegraphics{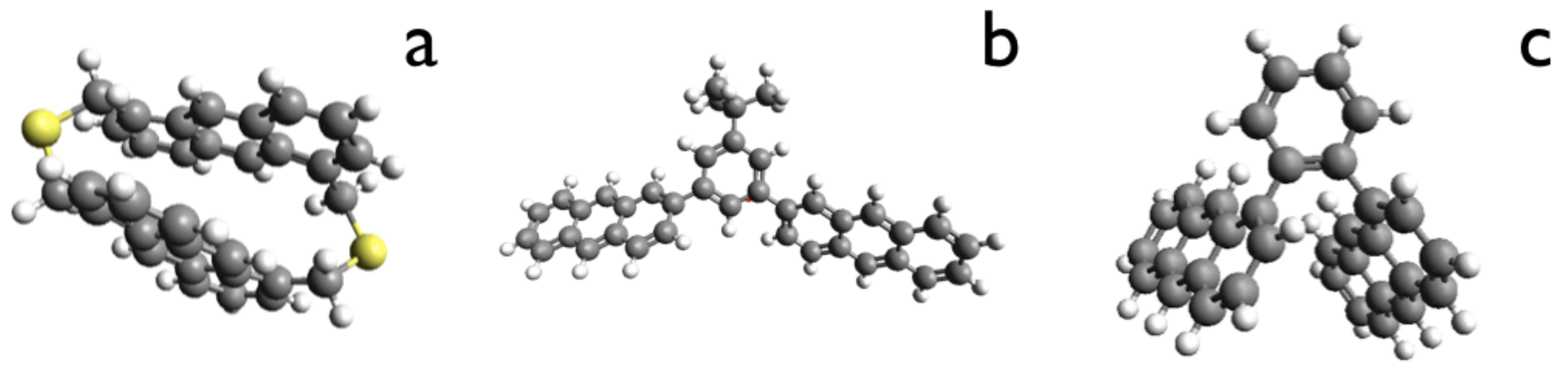}}
\caption{Excitonic dimers considered in this study, from left-to-right [2,2'] dithia-anthracenophane (DTA), 3,5-bis(anthracen-2-yl)-tert-butylbenzene (MDAB) and 1,2-bis-(anthracen-9-yl)benzene (ODAB).}
\label{fig:dimers}
\end{center}
\end{figure}

\indent 	Recent experimental evidence of coherent electronic energy transport at biologically relevant physical scales has spurred studies of the basic dynamics\cite{Dawlaty:2011ys}, and new methods for propagating the quantum system state\cite{Ishizaki:2009dc,Ishizaki:2009jg,Zhu:2011kx} as it interacts with the environment we cannot fully characterize. However detailed pictures of how the quantum state couples to the bath are usually absent from these studies, which often assume an environment characterized by a small number of parameters (for example a spectral density of the Drude-Lorentz form\cite{Ritschel:gb}). Models for the electronic coupling usually vary in detail\cite{Scholes:2003kx,Beenken:2004hc,Hsu:2009fk,Guthmuller:2009uq,Renger:2009uq,Munoz-Losa:2008fk} between the dipole approximation and the single-particle Coulomb interaction, but quantum many-particle effects are assumed to be negligibly small. It is not clear from the literature\cite{Sagvolden:2009fk} at present day how well electronic structure theory can provide all the parameters which are required to produce the timescale of coherence decay. To address this we focus on excitonic dimers\cite{Zewail:1975pi} as prototypes of coherent transport. At the outset the task would seem ambitious since the couplings between chromophores are typically on the order of 300cm$^{-1} \approx 1$ kcal/mol. Excited state methods which can be afforded for molecules of this size, mainly time-dependent density functional theory(TDDFT), are routinely in error\cite{Peach:2008dq,Schreiber:2008uq} by 1600cm$^{-1}$. This, and the large size of the systems of interest, are likely the underlying reason why atomistic calculations of decoherence dynamics are so infrequent in the literature. \\
\indent 	Recently an experiment\cite{Yamazaki:2003hd} has characterized coherent resonant energy transfer (CRET) in a closely related family of anthracene dimers (\ref{fig:dimers}). In this investigation an exciting femtosecond pulse of linearly polarized light prepares a superposition of electronic states on a pair of identical chromophores. Then over the course of about a picosecond, the time-dependent fluorescence is detected by upconversion with polarization sensitivity. The oscillations of the fluorescence anisotropy ($r(t)$) measured in these experiments are a marker of coherent transport (as demonstrated in the pioneering work of Hochstrasser\cite{Zhu:1993vn} and coworkers). The three dimers, [2,2'] dithia-anthracenophane (DTA), 3,5-bis(anthracen-2-yl)-tert-butylbenzene (MDAB) and 1,2-bis-(anthracen-9-yl)benzene (ODAB), are all derivatives of anthracene. Theoretically DTA has been examined\cite{Kishi:2009vn} by the groups of Cina\cite{Matro:1995ys,Biggs:2009fk} and Jang\cite{Yang:2010uq}, who supported the picture of coherent transport. The latter work provided direct calculations of the Coulomb coupling within a model-Hamiltonian picture of transport, and used DTA as a test-bed for an efficient new theory of CRET\cite{Jang:2008ef,Jang:2011uq}. Building on this previous work, this paper elucidates how well electronic structure can provide an ab-inito master equation for this phenomenon without assumption of site localized states, or classical coulomb($J$) coupling.\\ 
\indent 	A perturbative master equation approach is one tool in a set of complementary approaches to the exciton transport problem where nuclear (phonon) motion is the dominant bath. The main advantage of the master equation approach is it provides a reduced system density matrix, and only requires perturbative information provided by derivatives of system matrix elements and models of bath correlation functions. It should be viewed as the most affordable approach, and is feasible whenever the closed electronic dynamics can also be propagated, and the bath correlation function is known. Sophisticated and sometimes formally-exact master equations which incorporate non-Markovian effects through auxiliary density matrices that propagate along with the system\cite{Tanimura:1989ys,Ishizaki:2005vn,Shi:2009zr,Ishizaki:2009jg,Meier:1999fk} are a difficult starting point for atomistic simulations because the cost which is already significant with model-hamiltonians increases further in a non-linear way with the structure of the bath correlation function and number of system modes. If the bath correlation function cannot be approximated by a harmonic model\cite{Gerdts:1998fk} and requires nuclear dynamics\cite{Habenicht:2007uq,Shim:mb,Olbrich:2011cq}, the situation becomes rapidly more difficult. Moving towards increased accuracy and increased cost, there are several dynamical simulation techniques which can propagate electrons coupled to nuclei\cite{Iyengar:2005ys,Chapman:2011vn}, such as: surface hopping (SH)\cite{Tully:1971tg,Craig:2005zr,Mitric:2009fu}, Ehrenfest dynamics\cite{Li:2005jr,Kamisaka:2006fk,Margulis:1999ys}, and semi-classical approaches\cite{Tao:2010kl,Ceotto:2009vn}. These suffer from some difficulties of their own; the latter two do not preserve detailed balance without modification, and the former is not an acceptable model for coherence dephasing without some additions\cite{Subotnik:2011oq} which are topics of current study. At the peak of accuracy there are formally exact path-integral based approaches\cite{Makri:1995uq} which have provided vital insight into open quantum systems, but are a difficult starting point for complex simulations.\\
\indent 	In this work we employ the Redfield\cite{Redfield:1957fv,Redfield:1965dz} master equation. We will see below that \emph{within our simulations} the couplings between states are so much larger than the reorganization energies that Redfield theory is appropriate for this application\cite{Kjellberg:2006cl} and can furnish good results. However it is a perturbative model and fails entirely when the electronic couplings are small relative to the bath strength. For this reason it has been refined by more sophisticated, non-Markovian models which can replace it with little additional cost. In particular the polaron transformed theory\cite{McCutcheon:2011kx} of coherent transport\cite{Jang:2008ef,Jang:2011uq} is an attractive starting point for a master equation which hybridizes a correlated electron Liouville equation with a rigorous bath model. This direction should be pursed in future research. \\
\section{Theory}
\begin{figure}
\begin{center}
\resizebox{110mm}{!}{\includegraphics{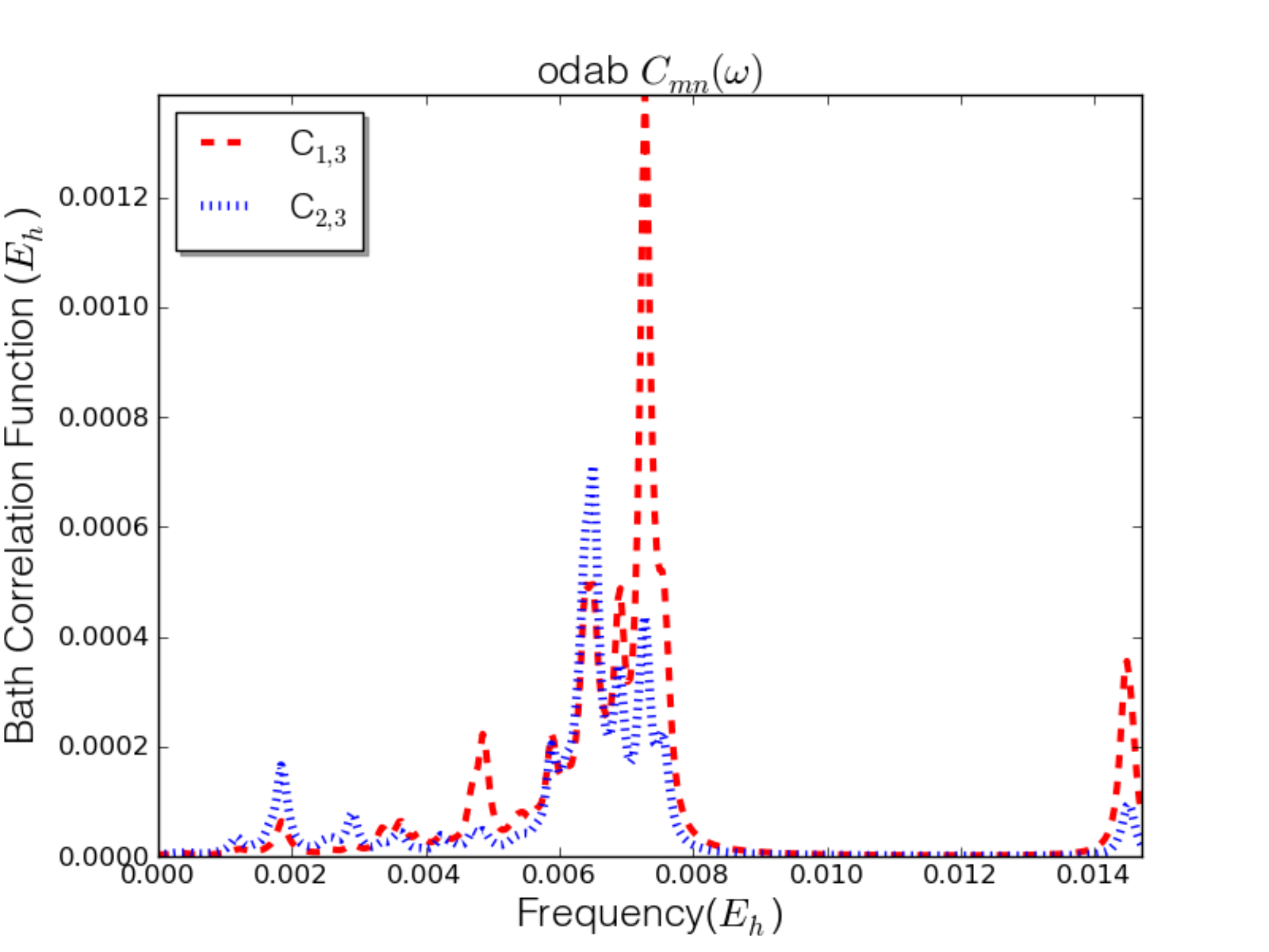}}
\caption{An example of the bath correlation functions produced from the approximations of this paper. $C_{mn}(\omega)$ is constructed for every pair of states below 6.5eV within the IMDHO model, and used to construct the Redfield tensor.}
\label{fig:cfunc}
\end{center}
\end{figure}

\begin{figure}
\begin{center}
\resizebox{140mm}{!}{\includegraphics{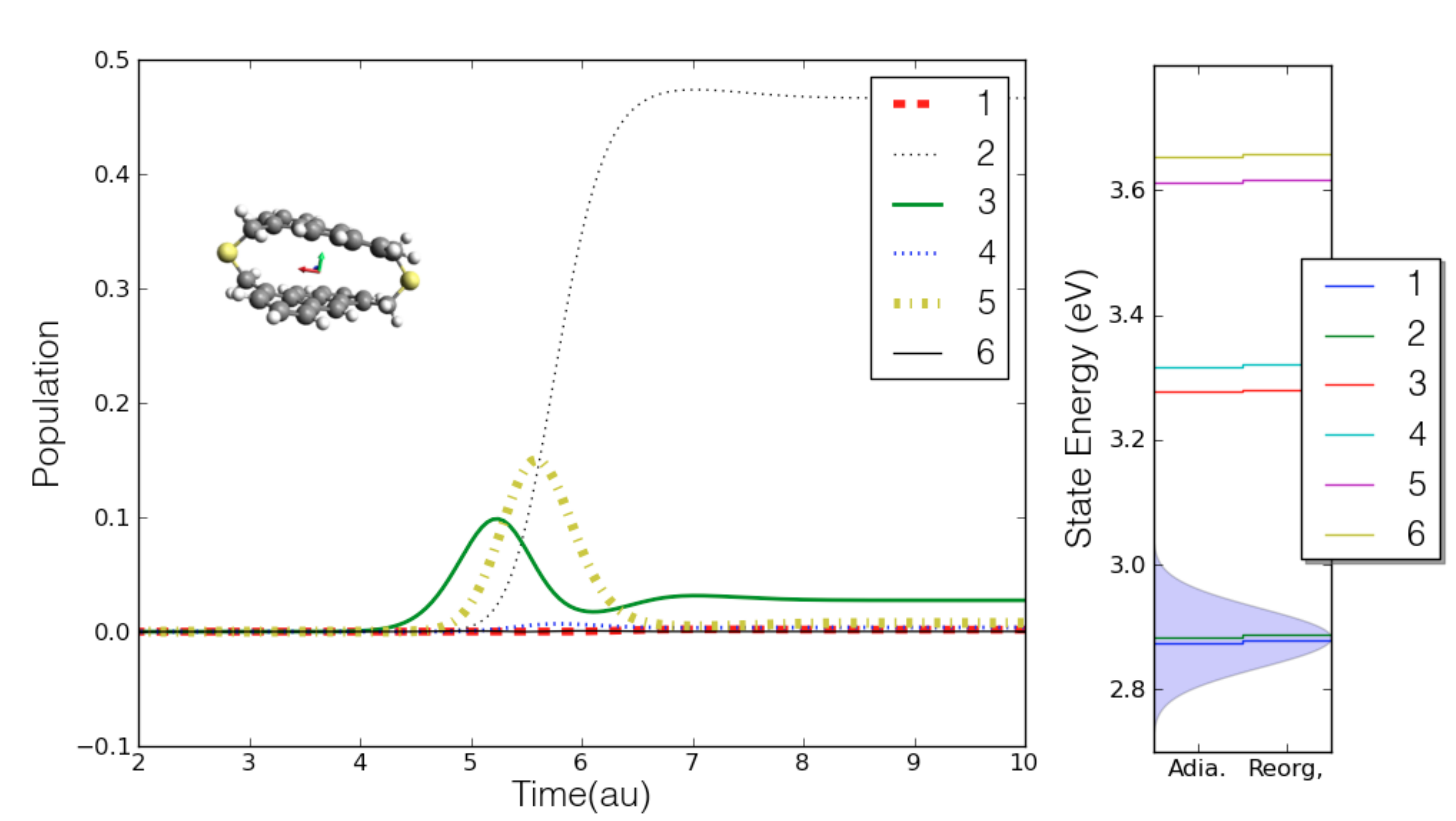}}
\caption{An example of how the [2,2'] dithia-anthracenophane(DTA) state populations evolve under the application of an 80fs oscillating electric field of 0.05 au along one of three axes (denoted by the red arrow in the figure). On the right, the corresponding stick spectra are shown, with and without reorganization energy, and in blue the envelope of the exciting pulse.}
\label{fig:pops}
\end{center}
\end{figure}

\indent	We wish to reproduce the decaying oscillations of the fluorescence anisotropy. The starting point of our analysis is the usual Liouville-Von Neumann equation\footnote[1]{Atomic units and the summation convention are assumed in all equations of this paper.} for the full unitary evolution of a molecule linearly coupled to a harmonic phonon bath,
\begin{align}
\dot{\sigma} = i\mathcal{L}\sigma = i[\hat{H},\sigma], \text{ where:} \hspace{1cm} \hat{H} = \hat{H}_\text{elec} +\hat{H}_\text{ph} + \hat{H}_\text{elec-ph}
\label{eq:liou}
\end{align}
and $\sigma$ is the full density matrix describing both electronic and nuclear degrees of freedom. We now introduce fermionic operators $a_i$ and $a^{\dagger}_i$, which respectively destroy and create an electron in an arbitrary many-electron state i (the exact nature of the electronic states is discussed below). Different choices yield different dimensionless displacements $ d_{i\alpha}$. Similarly, we introduce bosonic operators $b_\alpha$ and $b^\dagger_\alpha$, which respectively destroy and create a vibrational excitation in mode $\alpha$. We expand the terms in the molecular Hamiltonian and density matrix in terms of these basis states as,
\begin{eqnarray}
&& \hat{H}_\text{elec} = (h_{ij}  + \delta_{ij} \omega_\alpha d_{i\alpha}^2/2)a^{\dagger}_j a_i, \label{electronic} \\ && \hat{H}_\text{ph} = \omega_\alpha(b^\dagger_\alpha b_\alpha + \frac{1}{2}), \\ && \hat{H}_\text{elec-ph} = \omega_\alpha d_{i\alpha}a^{\dagger}_i a_i(b^\dagger_\alpha + b_\alpha), \\&& \sigma = \rho_{ij}a^{\dagger}_i a_j  \times \rho_{\alpha\beta}b^{\dagger}_\alpha b_\beta.
\end{eqnarray}
Here, $d_{i\alpha}$ is a dimensionless parameter measuring the strength of coupling of the ith electronic state to vibrational mode $\alpha$. Atomic units are used throughout this paper and it is assumed that repeated indices are to be summed over. In Eq. (2), we see that in addition to the usual matrix elements $h_{ij}$ of the electronic Hamiltonian, $\hat{H}_\text{elec}$ includes an additional contribution from reorganization energies of the bath, $\omega_\alpha d_{i\alpha}^2/2$. \\
\indent 	In order to simulate CRET,~\ref{eq:liou} must be integrated for a few picoseconds. It is not yet feasible\cite{Liang:2011ly,Akama:2010ly} to perform all-electron propagations for systems of this size over periods of time which are so long on the electronic timescale. Instead, we will treat the nuclear dephasing using a master equation, and propagate only an electronic density matrix.  We propagate the electronic density matrix using TDDFT within the Tamm-Dancoff approximation (TDA)\cite{Hirata:1999kh}, which obeys an equation of motion for an auxiliary one-particle density matrix which is the same as the electronic part of \ref{eq:liou}. Without the added approximation of the TDA, TDDFT (and TDHF) are non-linear, because of the dependence of the Fock operator on the state. As a result, resonant coherent Rabi oscillations don't properly appear in the adiabatic approximation, as was recently realized\cite{Ruggenthaler:2009zr}.\\
\indent	\emph{In principle, properties beyond the electron density can be provided with the addition of new functionals.} For the purposes of this work we make the assumption that the Kohn-Sham density matrix is a reasonable approximation to the true density matrix. This is a necessary first approximation to make headway, and can be relaxed by employing dynamical models which do rigorously provide the 1RDM\cite{Giesbertz:2009qa}. We make a small-matrix approximation to a Kohn-Sham-Redfield scheme to evaluate the usefulness of a harmonic bath model. In the Kohn-Sham-Redfield\cite{Tempel:2011uq} scheme, the Kohn-Sham single-particle density matrix $\rho^{ks}$ is propagated according to a Redfield master equation (see below). We choose the TDA from the outset, and assume that an approximate propagation spanned by the space of adiabatic stationary states ($\psi_i$) coming from TDDFT/TDA with energies below 6eV (the sum of excitation energy and pulsewidth) is sufficient to represent the combined dynamics. Coupling between the states will occur via the dipole operator during excitation and the Redfield relaxation operator described below. Transition moments between states evaluated at this level of detail\cite{Yeager:1975ve}, add a time-dependent off-diagonal term to $\hat{H}_{elec}$, $\hat{V}_i^j(t) = \vec{E}(t)\cdot\mu_{ij}$. With a more realistic density matrix than the Kohn-Sham density matrix these moments between states would change most significantly.\\
\indent 	At this point it's convenient to collect the time independent pieces of $\hat{H}$ and work in a basis which diagonalizes this part of $\hat{H}$. To this approximate electronic Liouville equation from electronic structure, we add dissipative terms\cite{Leathers:2006fk,Micha:1999kl} of a Markovian Redfield equation, affording an effective dynamics of the reduced system density matrix in the presence of the bath. The resulting Liouville equation is: 
\begin{align}
\dot{\rho}_{IJ} = i (\hat{H}_{IK}\rho_{KJ}-\rho_{IK}\hat{H}_{KJ}) + \mathcal{R}_{IJKL}(t) \rho_{KL}(t)  \\ 
\mathcal{R}_{IJKL} = \Gamma_{LJIK} + \Gamma_{KIJL}^* - \delta_{JL} \Gamma_{IMMK} - \delta_{IK}\Gamma_{JMML}  \\
\Gamma_{IJKL}(t) = \sum_{\alpha, i, j}  \int_0^\infty dt e^{i \omega_{KL} t}  C^\alpha_{ij}(t) \langle I | a^\dagger_{i}a_{i} |J \rangle \langle K| a^\dagger_{j}a_{j} |L \rangle 
\label{eq:red}
\end{align}
Indices $I,J$ have been capitalized to reflect the possibility that the basis of \ref{eq:red} may chosen differently than the basis of \ref{eq:liou}. For the purposes of electronic energy transport in molecules, the classic description of the bath, following Huang\cite{Born:1998ff}, is the vibrations which underly the molecule itself. To provide an independent mode displaced harmonic oscillator model (IMDHO) for the correlation function of these bath modes ($\alpha,\beta..$), a nuclear Hessian calculation is performed. $C^{\alpha}_{i, j}$ is the thermal equilibrium correlation function $\langle (d^\alpha_{i} \omega_\alpha \hat{x}_\alpha(t)) (d^\beta_{j} \omega_\beta \hat{x}_\beta(0) ) \rangle $. In the IMDHO model this only contributes for $\alpha = \beta$. The coupling constant between the states is then the so-called Huang-Rhys factor. Assuming that the potential surfaces of the lower and upper states are harmonic wells of the same frequency\cite{Petrenko:2007hx}, this coupling constant is given by the gradient of the diagonal electronic matrix element projected onto the mode and the mode frequency via: 
\begin{align}
S_{i\alpha} = \frac{\omega_\alpha d_{i\alpha}^2}{2} \textrm{ where }  d_{i\alpha}= \omega_\alpha^{-1} \frac{\delta \langle i | \hat{H}  | i\rangle}{\delta Q_\alpha} \\
d_{i\alpha}= \sum_{r_n}  \frac{\delta \langle i | \hat{H}  | i\rangle}{\delta r_n} U^\alpha_n (m_n\sqrt{\omega_\alpha})^{-1/2}
\label{eq:Huang}
\end{align}
where $d_{i\alpha}$ is the dimensionless displacement, $n$ is an atom with mass $m_n$, and $U^\alpha_n$ is the cartesian mass-weighted normal mode coordinate. The correlation function of a single nuclear mode is then given by the usual\cite{Mukamel:1995zt} expression:
\begin{align}
C^{\alpha}_{ij}(t) = \omega_\alpha^2 \frac{d_{i\alpha} d_{j\alpha}}{2} [ (\bar{n}_\alpha+1)e^{-i\omega_\alpha t}+(\bar{n}_\alpha)e^{i\omega_\alpha t}  ] \text{ where: } \bar{n}_\alpha = (e^{(\beta \hbar \omega_\alpha)}-1)^{-1}
\label{eq:corrfcn}
\end{align}
In a solution of tetrachloroethylene, the redistribution of vibrational quanta in anthracene has been studied at room temperature, with time-resolved Raman techniques and found to occur on timescales between 1-10ps\cite{Gottfried:1983uq}, which would correspond to a homogeneous vibrational linewidth of 5cm$^{-1}$ for one pair of modes. In addition to all the approximations above, we assume a Lorentzian broadening of the harmonic thermal correlation function by $\Gamma = 40cm^{-1}$, due to anharmonicity and collisions with the surrounding medium. Diabatic results are relatively insensitive to choices of this parameter (pure-dephasing lifetimes vary by less than a factor of two for reasonable values). The BLYP/6-31g** pure-dephasing lifetimes of of the bright states in DTA and MDAB are .894 and .566 ps with $\Gamma = 40$cm$^{-1}$, for $\Gamma = 100$cm$^{-1}$ the same two numbers are .900 and .564 ps. The result of this series of approximations are correlation functions between each state which resemble the sum of Lorentzian peaks in \ref{fig:cfunc}. These functions are different for every pair of states. In the direction of further rigor, correlation functions can be collected from classical or semi-classical dynamics calculations, or zero-frequency pure-dephasing contributions may be calculated from anharmonicities\cite{May:2004bs} of the excited state surface.\\
\indent 	The experimental observable we seek to reproduce is the fluorescence anisotropy $r(t) = (I_\parallel(t) - I_\perp(t)) /( I_\parallel(t) + 2 I_\perp(t)) $. Where $I_\parallel$ the fluorescence intensity parallel to the stimulation polarization. Factoring out the transition moments of the detector, and making the rotating wave approximations, fluorescence intensity is given\cite{Deng:1984ly} by the dipole dipole autocorrelation function: $I_{\gamma\delta}(t) \propto Tr(\hat{\rho}_0 \hat{\mu}_\gamma e^{i \mathcal{L} t} \hat{\mu}_\delta )$ where $\rho_0 = |0\rangle \langle 0|$ is the ground state and $\gamma, \delta$ cartesian indices. To spherical average in a numerical propagation of the Redfield equation, we apply an 80fs pulse to three orthogonal directions of the molecule with a carrier frequency that is the average energy of the first four excited states, resulting in a coherent superposition of populated states \ref{fig:pops}. The polarization which results in all three directions over time is then used to create a $I_{\gamma\delta}(t)$. The anisotropy is then evaluated at each time using the usual formulas\cite{Zare:1988bh}. $r(t)=(1-\rho_{dp})/(1+2\rho_{dp})$, where the depolarization ratio $\rho_{dp}$ is determined from the isotropic and anisotropic tensor invariants of $I_{\gamma\delta}$. The fit of Yamazaki's data with the functional form of Hochstrasser: 
\begin{align}
r(t) = \frac{0.1}{1+e^{-\frac{2 t}{T_2^{'}}} C} \left( (1+3C)+3(1-C)e^{-\frac{t}{T_2^{'}} }*Cos(\omega_{osc}t + \delta) + (3+C)e^{-\frac{2t}{T_2^{'}} } \right) \notag \\  \text { where } C = Cos^2\theta \text{ and }  \omega_{osc} = 4\beta^2 - (T_2^{'})^{-2}
\label{eq:fit}
\end{align}
is used to represent the experiment in this work. Since we use a many-state model we interpret $\beta$, and $T_2^{'}$ as \emph{effective} couplings and dephasing times respectively. $\theta$ is related to the angle between transition moments, and $\delta$ is a phase-shift.
\subsection{Choice of basis and solvation}
\indent 	As in the perturbation theory of the electron-correlation problem\cite{Lochan:2007fk}, the choice of basis states used in Eq. \ref{eq:Huang} has a significant impact on the results of master equations. This is well-appreciated in the polaron-transform approach of Silbey and coworkers\cite{Rackovsky:1973ve}. In most applications of master equations to excitonic systems, a somewhat local basis is assumed\cite{Holstein:1959rq}, in which $\hat{H}_{elec}$ is not diagonal, and the system-bath coupling is also only assumed to occur only on the diagonal elements of the density matrix. These diagonal bath couplings are rotated off the diagonal when the zeroth-order Hamiltonian is formed (which includes electronic and bath reorganization contributions) and thus provide relaxation between the stationary states when the dynamics is performed. This model which resembles the Holstein hamiltonian\cite{Holstein:1959rq} does not emerge directly from electronic structure theory. The question becomes how to define an atomistic prescription for $\mathcal{R}$ without resorting to dynamics.\\
\indent 	If one naively choses the excited states of a single electronic structure calculation to prepare a master equation, ie: $h_{ij} = 0$ $(\forall i \neq j) $, there are several problems. In the Markovian case adiabatic state-derivatives only contribute to $\mathcal{R}$ via $C(\omega)$ at $\omega = 0$, where there are no physical vibrational peaks on which the gradient can be projected in Eq.(\ref{eq:Huang}). Rigorous pure-dephasing contributions at this frequency can be related to anharmonicities\cite{May:2004bs} which are too expensive to calculate for systems of this size. If fluctuations of non-adiabatic couplings $d_{mn}(R) = \langle m|\frac{d}{dR'}|n\rangle |_{(R' = 0)}$ under the equilibrated bath were included this would no longer be the case. The reorganization energy corresponding to $d_{mn}$ would also result in mixed zeroth-order states, which would to some extent be localized. However the success of a gradient-based IMDHO model for the spectral density of a non-adiabatic coupling would not be very good, because the non-adiabatic coupling's shape\cite{Valero:2008uq} is very far from that of a harmonic oscillator, and near zero at equilibrium. One would need to resort to optimizing this quantity, or performing dynamics to determine a dimensionless displacement. Moreover there is an issue that the couplings \emph{between excited states} are second order response properties of the TDDFT Lagrangian and have not yet been reported for DFT\cite{Send:2010bu}.\\
\indent 	In an entirely detailed picture, a reorganization Hamiltonian for solvent polarization degrees of freedom which are very slow on the electronic timescale, and have little impact on the dynamics, would localize the electronic states of the system. One effective way to incorporate these effects into our simulation is to adopt a recently implemented diabatization\cite{Subotnik:2009ck,Vura-Weis:2010kx} procedure invented by Subotnik and coworkers which mixes the CIS (or TDA) states with one-another such that their interaction with an implied dielectric is minimized. Again the electron-phonon coupling is taken to be diagonal in the electronic basis, although the Hamiltonian now has off-diagonal Coulomb and exchange (if it is present in the density functional)  coupling. An admitted drawback of this procedure is that the poles of the $0^{th}$ order Hamiltonian are not shifted by their interaction with the solvent model, however the ease of calculation and the results we will obtain strongly support the usefulness of this idea. Because the gradients of the diagonal elements of these states are no longer analytically available, we have evaluated the gradient of the diabatic Hamiltonian provided by this procedure by central differences.
%
%
%
\section{Results}
\subsection{Stationary parameters.}
\indent 	The vertical ($\omega$B97//6-31g**)\cite{Chai:2008nx} TDDFT excitation spectrum of a single anthracene molecule has two poles below 5eV at 4.28, and 4.43 eV, with only the lower state possessing significant oscillator strength. This is a reasonable reproduction\cite{Zilberg:1995fk} bright $^1$B$_{1u}$ and dark $^1$B$_{2u}$ states which occur experimentally at 3.43 and 3.47 eV respectively. For some of the dimers studied the stick spectrum which results can be rationalized as simple mixture of these states: with two bright states coming from symmetric and antisymmetric combinations of the bright states amongst themselves. The dark states, whose realism was questioned in previous work\cite{Yang:2010uq}, appear to have charge-transfer character, and their relative position to the bright states in the spectrum changes depending on whether an asymptotically correct exchange functional is used or not. The TDDFT stick spectra are in good qualitative agreement with a correlated wavefunction based\cite{Head-Gordon:1999ss,Rhee:2009fc} excited state theory (\ref{tbl:cisd0}). The low lying dark states are likely real, and indeed in our dynamics simulations they are involved in population relaxation of the bright states.\\
\begin{table}[h!]
\begin{tabular}{c c c c c c }
\hline
Method & Species & State 1 & State 2 & State 3 & State 4 \\
\hline
RI-SOS-CIS(D$_0$)&DTA&	3.83, f=0.077 &	3.84, f=0.047	& 3.89, f=0.104 &	3.89, f=0.026	\\
&MDAB& 	3.89, f=0.003 &	3.90, f=0.004	& 4.00, f=0.150 &	4.01, f=0.175	\\
&ODAB &	3.93, f=0.023 & 3.95, f=0.001 &	3.98, f=0.158	& 4.05, f=0.266 	\\
\hline
B3LYP&DTA&	2.84 , f=0.002 & 2.85, f=0.002 & 3.22, f=0.043 &	3.26, f=0.069\\
&MDAB& 	3.19, f=0.001 &	3.20, f=0.001	& 3.37, f=0.153 &	3.38, f=0.050	\\
&ODAB &	3.00, f=0.032 & 3.07, f=0.021 &	3.37, f=0.104	& 3.44, f=0.128 	\\
\end{tabular}
\caption{Singlet excited state energies (eV) and oscillator strengths, f, produced by RI-CIS(D$_0$)//cc-pvdz follow the same trends as the B3LYP results, although the values of the couplings are more poorly reproduced.}
\label{tbl:cisd0}
\end{table}
\indent 	The semi-quantitative agreement of electronic couplings ($\beta$) with the experimental oscillation periods is the most difficult aspect of the experiment to reproduce with TDDFT, although it was not the focus of this work. If the Hamiltonian is expressed in a basis of localized, degenerate states $\beta$ is the off diagonal element between two states. In the basis of states adiabatic which diagonalize the electronic Hamiltonian, it is related to half the gap between the energies of the bright states. The experimental estimate of $\beta$ from the oscillation period depends on a 2-state model, and the subtraction of a relaxation lifetime. Theoretical models depend on either perturbation theory of monomer densities which have been separated in an ad-hoc way, or a splitting between adiabatic energies. The latter, super-molecule approach is more realistic and can capture whatever electron interaction effects are present in the underlying electronic structure model (although the perfect mixture of the two states is assumed). Because of the uncontrolled nature of the approximations relating the experimental $\beta$ and electronic state gap, one should view direct comparison of the experimental and calculated $\beta$ semi-quantitatively. Only the final simulated oscillation should be directly comparable to the experiment. Still if the adiabatic state-splitting is more than an order of magnitude larger than the experimental $\beta$ no reasonable simulation will be possible, because the reorganization energies will not shift the oscillation into the correct regime.\\
\indent 	The previous work\cite{Yang:2010uq} calculated state couplings for DTA from energy differences, transition densities, and the dipole approximation, with several choices of density functional and geometry. Solvent effects on the coupling were also calculated (and found to be small $\approx$7cm$^{-1}$). A range of values between 44 and 144 cm$^{-1}$ was obtained, and the authors identified the agreement with the experimental values as somewhat fortuitous. We find that for these covalently bonded dimers, going up-to exchange effects is insufficient (\ref{tbl:coup}) to reproduce the qualitative trend in coupling energy. In \ref{tbl:coup} we calculate the coupling energy with increasingly sophisticated electronic structure models. The basis is held constant, but even in larger bases the conclusions remain the same (see \ref{tbl:basis}). In the first column the monomers interact mostly via a classical coulomb interaction because the electronic structure lacks long-range exchange. The qualitative trend is correct. In the second and third, more exchange is added and the relative agreement becomes worse. Finally with exchange and a second-order treatment of correlation, good relative agreement is obtained. It seems possible that coulomb only methods (TDC, BLYP, dipole-dipole, etc.) benefit from some cancellation of non-local exchange and correlation errors.\\
\indent 	These findings are in keeping with existing work\cite{Sagvolden:2009fk,Scholes:1993ly} which has found that Forster and Dexter coupling are insufficient for nearby molecules, and indeed in error up to a factor of two. It seems likely that for these systems separated by less than 5 Angstrom, long-range correlation effects are required to reproduce the correct trend of energy splittings. Solvation effects on related chromophore pairs have been studied more thoroughly than medium-range correlation, and are known to be meaningful\cite{Munoz-Losa:2008fk}. The roughly twenty percent reductions of the coupling relative to vacuum have been reported for naphthalene dimer in a continuum model of hexane. Ideally we would fully treat both effects, unfortunately, the costs of these calculations are prohibitive for pursuing them as the basis for the decoherence rate. \\
\indent 	We do not challenge the validity of the transition dipole, or transition density cube\cite{Madjet:2006bd} approximations when applied to distant chromophores. Rather we are suggesting that in cases of nearby chromophores where long-range correlation would be considered a prerequisite for ground state thermochemistry\cite{Grimme:2006qc}, it's likely also necessary to reproduce excitonic couplings\cite{Liu:2011kx}. Correlated wave-function based models, or double hybrids should be applied to these problems if possible. Simply choosing experimental couplings, would be inappropriate since the basis of adiabatic electronic states from which we calculate $C(t)$, and the experimentally accessed states are different. This approach would also have difficulty accessing the dark charge-transfer states. Phenomenological approaches which combine experimental energies with experimental bath parameters are consistent and successful. However combining calculated electronic spectra with experimental bath correlation functions ignores the difference between the basis states of the experiment and the calculation. \\
\begin{table}[h]
\centering
\begin{tabular}{ c c c c c c }
\hline
	& BLYP & B3LYP & $\omega$B97 & RI-SOS-CIS(D)$_0$  & Exp. \\
\hline
DTA &64	& 150 &	220 &	4 &	14 \\
MDAB& 88	& 84 &	20 &	18 &	29 \\
ODAB& 266	& 284 &	429 &	273 &	51 \\
\end{tabular}
\caption{ State couplings $\beta$(cm$^{-1}$) as estimated from half the difference of bright excited state energies in the cc-pvdz basis for various methods differing in their treatment of exchange and correlation. The agreement is best with the addition of long-range correlation (last column). The addition of exchange appears to worsen the results. Basis set dependence is addressed below.}
\label{tbl:coup}
\end{table}
\begin{table}[h]
\centering
\begin{tabular}{ c c c }
\hline
	& B3LYP & BLYP \\
\hline
DTA & 144 & 93 \\
MDAB& 90	& 80 \\
ODAB& 257 & 246 \\
\end{tabular}
\caption{ State couplings $\beta$(cm$^{-1}$) as estimated from half the difference of bright excited state energies in the QZVP basis. With this basis these quantities are nearly converged with respect to basis-set size, but TDDFT is still unable to reproduce the qualitative trend.}
\label{tbl:basis}
\end{table}
%

%
%
\section{Difficulty of an adiabatic approach.}
\begin{figure}[h]
\begin{center}$
\begin{array}{cc}
\resizebox{70mm}{!}{\includegraphics{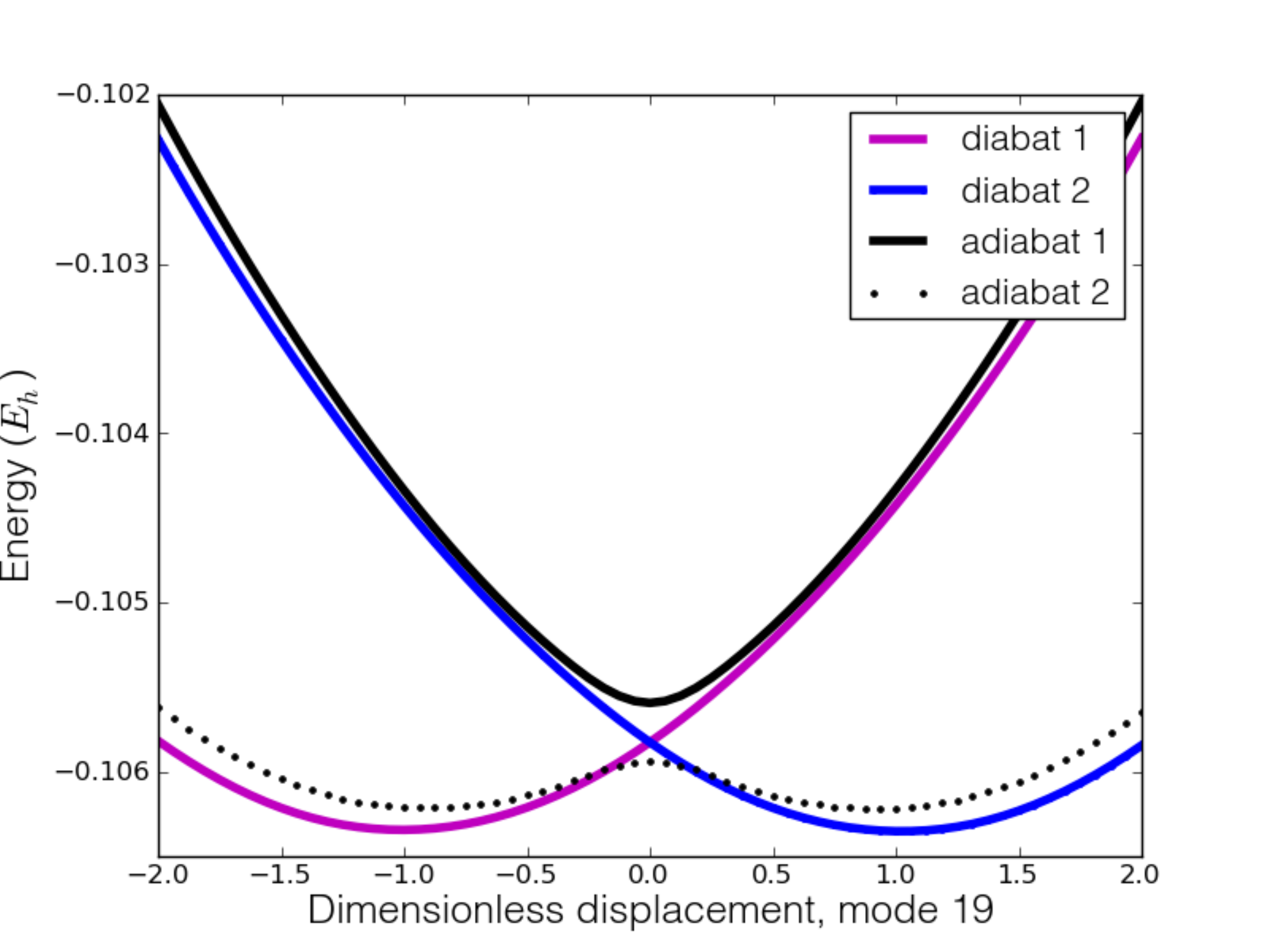}} & 
\resizebox{70mm}{!}{\includegraphics{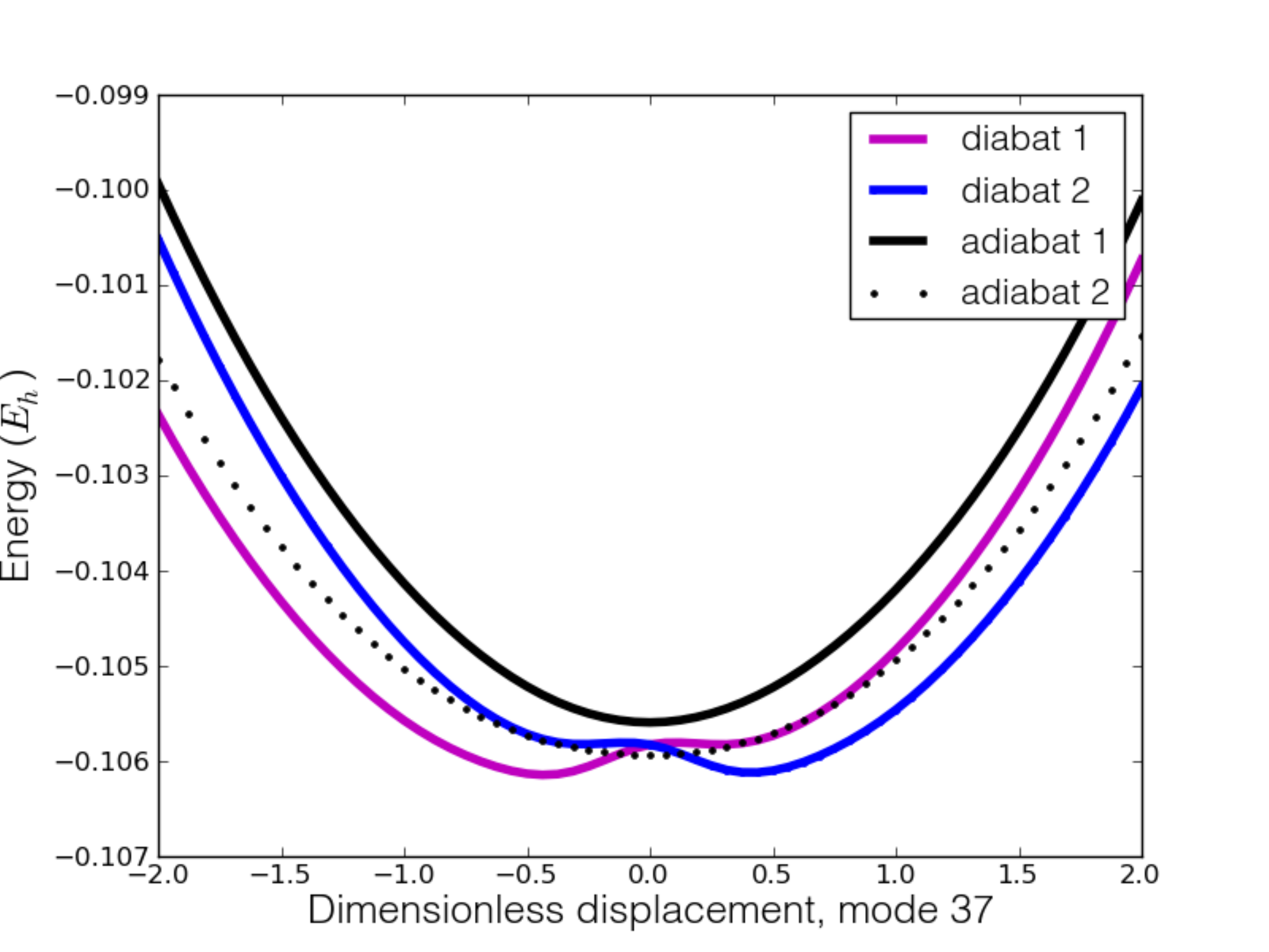}}
\end{array}$
\caption{Energies of adiabatic, and quasidiabatic transfer states as a function of dimensionless displacement along [2,2'] dithia-anthracenophane's 19$^{th}$(left), and 37$^{th}$(right) normal mode. In the figure on the left it's immediately clear by inspection that the diabatic states are more harmonic than the adiabat's. The figure on the right is an example where the accuracy of using Eqn. \ref{eq:corrfcn} is questionable.}
\label{fig:cob}
\end{center}
\end{figure}

\indent	We have already outlined several difficulties of using adiabatic states in a Kohn-Sham Redfield scheme. Within the Markovian approximation bath correlation only appears at zero-frequency. Another problem which afflicts both choices of basis to some degree is the anharmonic structure of the bath correlation function. It is always possible to write a linear system-bath coupling in terms of HO's\cite{Feynman:1963kx}, but the appropriateness of the IMDHO, or even an oscillator model with frequency changes for a dimer system depends on the strength of the vibronic interaction \cite{Beenken:2002il}. The adiabatic states often do not resemble displaced harmonic oscillators (\ref{fig:cob}), but rather possess a double-well type structure, which means that even a proper Markovian master equation would require dynamics to obtain the bath correlation function. In the case of most modes, this is repaired within the quasi-diabatic states. Some other diabatization\cite{Pacher:2007vn} procedures directly mix states to eliminate electron nuclear coupling, and along these lines one could construct a basis which leaves the electronic surfaces optimally harmonic which would be useful for a rudimentary model of nuclear motion. In any case the coherence lifetimes produced (\ref{tbl:lifetimes}) from our simulations based on adiabatic states are in much poorer agreement with the experimental lifetimes than those using the procedure of Subotnik and co-workers, as discussed further below. Because only the zero-frequency part of the correlation function is meaningful for work based on an adiabatic basis. 
\begin{table}[h]
\centering
\begin{tabular}{ c c c c }
\hline
&&		Adiabatic	&Diabatic\\
\hline
Species & Exp.	&B3LYP/TZVP& 	B3LYP/6-31g*\\
\hline
DTA &	1.20&	2.14	&1.40 \\
MDAB &	0.73	&0.03&	1.80\\
ODAB & 1.02	& 0.12&	1.39\\
\end{tabular}
\caption{ Pure-dephasing time (ps) between the two brightest states ($1/\mathcal{R}_{mnmn}$), as obtained from two different choices of electronic basis, compared to the experimental $T_2^\text{'}$.}
\label{tbl:lifetimes}
\end{table}
\section{Quasi-Diabatic Approach.}
\begin{figure}
\begin{center}$
\begin{array}{cc}
\resizebox{70mm}{!}{\includegraphics{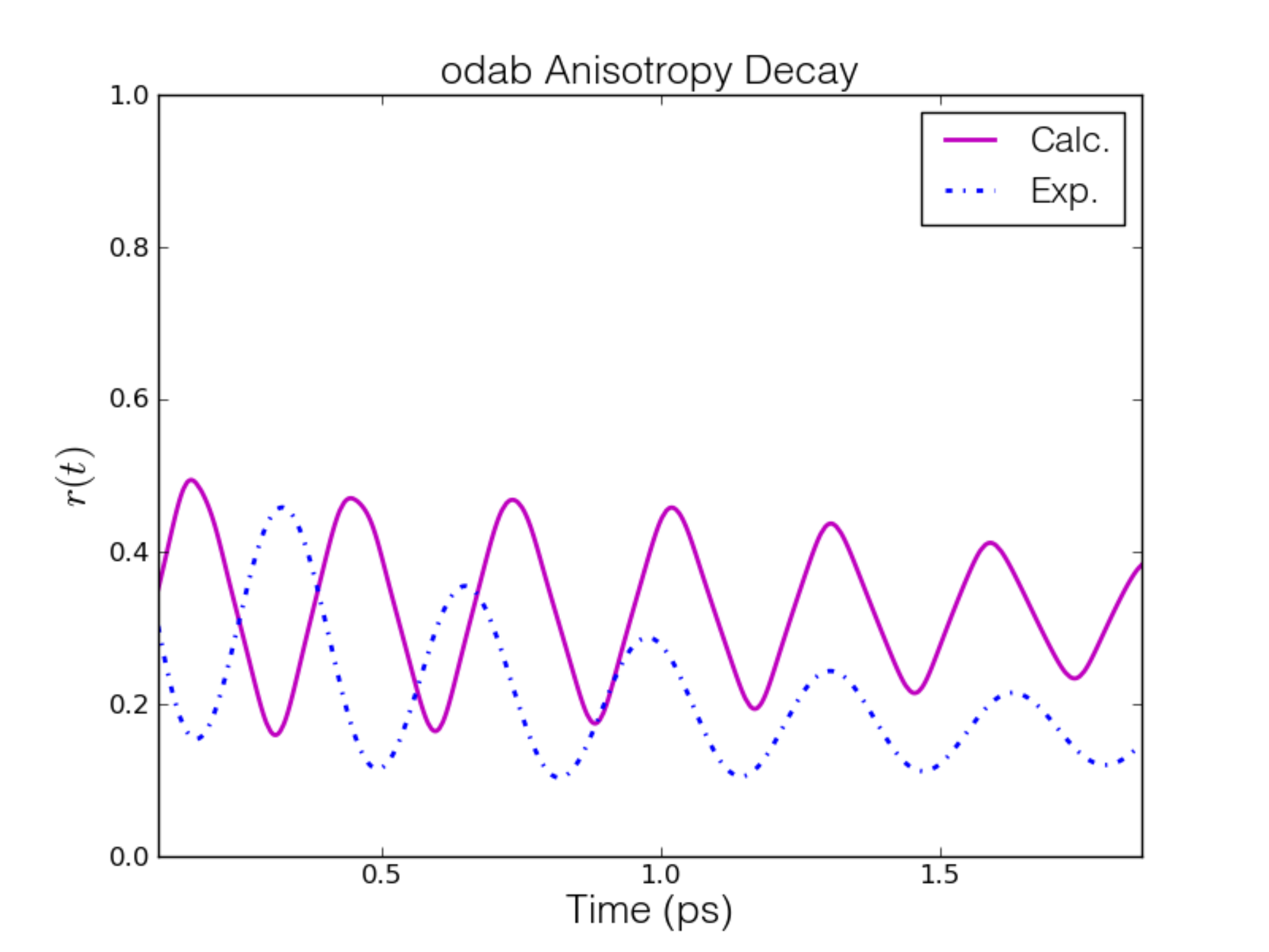}} & 
\resizebox{70mm}{!}{\includegraphics{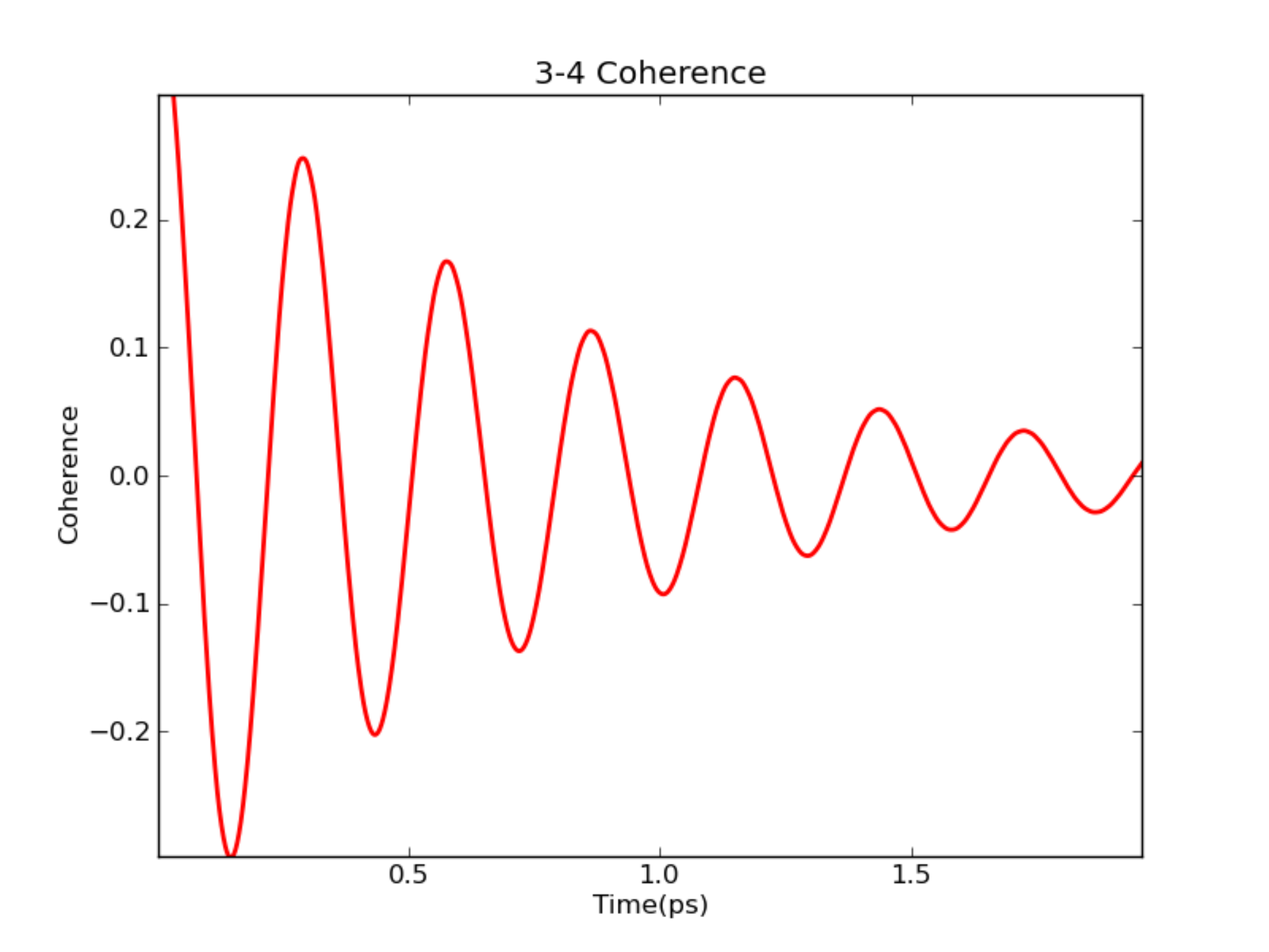}}
\end{array}$
\caption{(The polarization anisotropy decay of ODAB calculated in this work ($\omega$B97//6-31g**), and compared to the analytically fit experimental decay of Yamazaki. On the right the evolution of the real part of the coherence ($\rho_{34}(t)+\rho_{43}(t)$). The decay of the coherence between these two states is much faster than the decay of the observable.}
\label{fig:odabdecay}
\end{center}
\end{figure}

\indent 	The experiments\cite{Yamazaki:2002ta,Yamazaki:2003hd} provide noisy oscillating flourescence anisotropy curves, which were fit in that work to Hochstrasser's\cite{Zhu:1993vn} functional form. We have directly compared the polarization anisotropy decays produced in this work from first principles, to the experimental data's fit curve. When examining \ref{fig:odabdecay}, the reader should keep in mind the absence of any rotational relaxation in our simulated anisotropy, and allow for an arbitrary relative phase shift between the coherent oscillations of experiment and theory. A low-pass filter (a step function non-zero below .07$E_h$) has been applied to the calculated signal to remove high frequency components which might be averaged over by the detector response time. The same simulation has been performed (\ref{tbl:lifetimes}) for all three molecules, in the hope that the relative trend of the coherence lifetimes could be reproduced. Although with vibronic coupling emerging from a diabatic basis all the decoherence lifetimes are within a factor of two of their experimental counterparts, the delicate relative trend of the lifetime isn't reproduced. This isn't surprising given that the relative trend of couplings is also not reproduced. The anisotropy decay can be compared (\ref{fig:odabdecay}) with the evolution of the coherence between bright states, which largely correspond to the third and fourth states ($\rho_{34}(t)$). This is the picture of decoherence which may come from a less-detailed model of the dynamics. The much more rapid decay of the coherence element signals the breakdown of a two-state model picture of the anisotropy. The anisotropy decays of DTA and mDAB are not shown, because in these cases the oscillations of the anisotropy are more than a factor of 5 too fast (because the couplings are so severely overestimated by TDDFT). 
%
%
%
\section{Discussion and Conclusions}
\indent 	Cina's previous work\cite{Biggs:2009fk} introduced a phenomenological 2-state ($J$ = 22.9 cm$^{-1}$) model of DTA coupled explicitly to an intra-chromophore vibration ($\omega_{12}$ = 385cm$^{-1}$, d = .55). It was found that coherent vibrational dynamics could be used to modify the dynamics of excitation transfer, although only weakly in the case of DTA. In the theory of Jang and coworkers\cite{Yang:2010uq} the electronic coherence oscillation period is given by the bare coupling renormalized by a bath contribution which depends on the parameters of a super-ohmic spectral density. Reversing these parameters such that they reproduce the experimental vibrational progression and oscillation period, they estimate a bare (purely electronic) coupling for DTA between 53 and 100$cm^{-1}$, about 40cm$^{-1}$ less than their B3LYP calculations. In this work the ODAB fluorescence anisotropy decay, and the coherence lifetimes of MDAB and DTA are satisfactorily reproduced from a TDDFT/TDA based Redfield model of the electronic system. However the TDDFT/TDA is found to overestimate bare electronic couplings of these dimers, and misses the relative trend with many choices of basis and functional. Long-range correlation is implicated as another physical effect which reduces the bare Coulomb coupling.\\
\indent 	The theory of CRET is mature\cite{Jang:2008ef,Jang:2011uq} within its assumed model of the system and bath, but deriving those models from first-principles electronic structure calculations is not yet routine. It should be pursued further, because two-state, uncorrelated models of the electronic dynamics are now much less realistic than sophisticated\cite{Zhu:2011kx,Huo:2010ys} treatments of system-bath correlation. An accurate calculation of the coupling between chromophores separated by less than 5 Angstroms likely requires a treatment of long-range correlation, beyond the scope of local TDDFT. However with available technology, a \emph{single} calculation of excited energies for these molecules which treats long-range electron correlation is a somewhat demanding computation. To provide a fully predictive picture of coherent transport the cost of correlated electronic structure should be further reduced, and the CRET theories should be hybridized with a correlated, many-state Liouville equation. This paper suggests that a picture of the bath as broadened nuclear motion, where solvation serves to localize\cite{Subotnik:2009ck,Vura-Weis:2010kx} the electronic states provides reasonable qualitative predictions of the decoherence rate. Since our model of the ODAB fluorescence anisotropy emerges directly from calculated data 
\subsubsection{Computational Details}
\indent 	Structures of [2,2'] dithia-anthracenophane (DTA), meta- Dianthraceneophane(MDAB) and ortho-Dianthraceneophane(ODAB) were optimized using the B3LYP\cite{Becke:1993bh} functional and def2-TZVP basis\cite{Schafer:1994ly} (and auxiliary basis\cite{Weigend:2006tw}) in the ORCA\cite{Neese:2008cr} program package, the same model used to evaluate their harmonic vibrational frequencies and modes. These geometries and modes were re-used for excited state calculations. A local exchange approximation, COSX\cite{Neese:2009ye} was also invoked. Transition moments between states and excitation energies were obtained from the Q-Chem program package\cite{qchem30} evaluated as expectation values of the dipole operator with TDA density matrices using the functionals and basis sets described in the text. The Redfield equation was integrated with a 0.05 atomic unit timestep, via a basic 4$^{th}$ order Runge-Kutta propagation, after an 80fs gaussian-enveloped 0.05 atomic unit oscillating electric field was applied as a stimulating pulse. The harmonic correlation function was evaluated at 273.15K. 
\section{Acknowledgments}
This work was supported by the Defense Advanced Research Project Agency under Contract No .FA9550-08-1- 0285 and the Defense Threat Reduction Agency under Contract No. HDTRA1-10-1-0046. The authors thank Prof. Alex Einsfeldt, and Dr. Dmitrij Rappoport for valuable conversations. A.A.G. thanks the Camille and Henry Dreyfus and Sloan foundations for their generous support. \\
%

\bibliography{Fret}

\end{document}